\documentclass[%
 aip,
 amsmath,amssymb,
 reprint,%
]{revtex4-2}

\usepackage{graphicx}
\usepackage{dcolumn}
\usepackage{bm}

\usepackage[utf8]{inputenc}
\usepackage[T1]{fontenc}
\usepackage{mathptmx}
\usepackage{etoolbox}

\usepackage[normalem]{ulem}
\usepackage{xcolor}
\usepackage{float}
\usepackage{CJKutf8}
\usepackage{graphicx}
\usepackage[hidelinks]{hyperref}

\begin{document}

\title{Hysteretic responses of nanomechanical resonators based on crumpled few-layer graphene}

\author{Heng Lu}
\thanks{These authors contributed equally.}
\affiliation{School of Optoelectronic Science and Engineering \& Collaborative Innovation Center of Suzhou Nano Science and Technology, Soochow University, Suzhou 215006, China}
\affiliation{Key Lab of Advanced Optical Manufacturing Technologies of Jiangsu Province \& Key Lab of Modern Optical Technologies of Education Ministry of China, Soochow University, Suzhou 215006, China}

\author{Chen Yang}
\thanks{These authors contributed equally.}
\affiliation{School of Optoelectronic Science and Engineering \& Collaborative Innovation Center of Suzhou Nano Science and Technology, Soochow University, Suzhou 215006, China}
\affiliation{Key Lab of Advanced Optical Manufacturing Technologies of Jiangsu Province \& Key Lab of Modern Optical Technologies of Education Ministry of China, Soochow University, Suzhou 215006, China}

\author{Ce Zhang}
\affiliation{School of Optoelectronic Science and Engineering \& Collaborative Innovation Center of Suzhou Nano Science and Technology, Soochow University, Suzhou 215006, China}
\affiliation{Key Lab of Advanced Optical Manufacturing Technologies of Jiangsu Province \& Key Lab of Modern Optical Technologies of Education Ministry of China, Soochow University, Suzhou 215006, China}

\author{YuBin Zhang}
\affiliation{School of Optoelectronic Science and Engineering \& Collaborative Innovation Center of Suzhou Nano Science and Technology, Soochow University, Suzhou 215006, China}
\affiliation{Key Lab of Advanced Optical Manufacturing Technologies of Jiangsu Province \& Key Lab of Modern Optical Technologies of Education Ministry of China, Soochow University, Suzhou 215006, China}

\author{FengNan Chen}
\affiliation{School of Optoelectronic Science and Engineering \& Collaborative Innovation Center of Suzhou Nano Science and Technology, Soochow University, Suzhou 215006, China}
\affiliation{Key Lab of Advanced Optical Manufacturing Technologies of Jiangsu Province \& Key Lab of Modern Optical Technologies of Education Ministry of China, Soochow University, Suzhou 215006, China}

\author{Yue Ying}
\affiliation{CAS Key Laboratory of Quantum Information, University of Science and Technology of China, Hefei 230026, China}
\affiliation{Suzhou Institute for Advanced Research, University of Science and Technology of China, Suzhou 215123, China}

\author{Zhuo-Zhi Zhang}
\affiliation{CAS Key Laboratory of Quantum Information, University of Science and Technology of China, Hefei 230026, China}
\affiliation{Suzhou Institute for Advanced Research, University of Science and Technology of China, Suzhou 215123, China}

\author{Xiang-Xiang Song}
\affiliation{CAS Key Laboratory of Quantum Information, University of Science and Technology of China, Hefei 230026, China}
\affiliation{Suzhou Institute for Advanced Research, University of Science and Technology of China, Suzhou 215123, China}

\author{Guang-Wei Deng}
\affiliation{Institute of Fundamental and Frontier Sciences, University of Electronic Science and Technology of China, Chengdu 610054, China}

\author{Ying Yan}
\affiliation{School of Optoelectronic Science and Engineering \& Collaborative Innovation Center of Suzhou Nano Science and Technology, Soochow University, Suzhou 215006, China}
\affiliation{Key Lab of Advanced Optical Manufacturing Technologies of Jiangsu Province \& Key Lab of Modern Optical Technologies of Education Ministry of China, Soochow University, Suzhou 215006, China}

\author{Joel Moser}
\email{j.moser@suda.edu.cn}
\affiliation{School of Optoelectronic Science and Engineering \& Collaborative Innovation Center of Suzhou Nano Science and Technology, Soochow University, Suzhou 215006, China}
\affiliation{Key Lab of Advanced Optical Manufacturing Technologies of Jiangsu Province \& Key Lab of Modern Optical Technologies of Education Ministry of China, Soochow University, Suzhou 215006, China}

\begin{abstract}
Manipulating two-dimensional materials occasionally results in crumpled membranes. Their complicated morphologies feature an abundance of folds, creases and wrinkles that make each crumpled membrane unique. Here, we prepare four nanomechanical resonators based on crumpled membranes of few-layer graphene and measure their static response and the spectrum of their dynamic response. We tune both responses with a dc voltage applied between the membrane and an underlying gate electrode. Surprisingly, we find that all four resonators exhibit hysteretic responses as the gate voltage is increased and then decreased. Concomitant discontinuities in the static response and in the vibrational resonant frequencies indicate a sudden change in the shape and in the tensile strain of the membranes. We also find that the hystereses can be removed and regular responses can be restored by annealing the resonators. We hypothesize that the hysteretic nature of the responses may originate from an interplay between the rugged morphology of the membranes and adsorbates trapped within the confine of the folds.
\end{abstract}

\maketitle
\newpage

Hallmarks of nanomechanical resonators based on low-dimensional materials are the tunability of their static displacement \cite{Storch2018} and the tunability of their vibrational resonant frequencies \cite{Sazonova2004,Bunch2007,Chen2009,Ye2021,Xu2022}. The tuning knob is often a dc voltage $V_\mathrm{gate}$ applied between the resonator and an underlying gate electrode that bends the resonator and modifies its spring constant \cite{Sazonova2004,Carr1997,Lifshitz2008}. In the case of resonators based on two-dimensional (2-D) membranes, tuning the displacement allows one to optimize the transduction of the vibrational amplitude into measurable signals \cite{Storch2018}, to measure the Young's modulus of the membrane \cite{Storch2018,Nicholl2015}, and to adjust dissipation \cite{Song2012,Barton2012,Chen2018}. Tuning the resonant frequencies makes it possible to engineer vibrational energy pathways \cite{Guettinger2017,Mathew2016}, to parametrically amplify vibrations \cite{Mahboob2008,Eichler2011,Mathew2016,Su2021}, to detect weak forces \cite{Weber2016}, and to create nanomechanical filters \cite{Chen2009} and oscillators \cite{Chen2013}. In the absence of hysteretic processes such as structural phase transitions \cite{Chaste2020}, magnetostriction \cite{Jiang2020}, and coupling between vibrations and superconductivity \cite{Sahu2022}, the dependences of the static displacement and the resonant frequencies on $V_\mathrm{gate}$ are generally reversible and non-hysteretic.

Recently, resonant frequency hystereses (RFH) upon increasing and decreasing $\vert V_\mathrm{gate}\vert$ have been reported in 2-D resonators, and various mechanisms have been put forward to explain their existence. In van der Waals heterostructures, RFH have been proposed to originate from interlayer slips (Ref.~\cite{Kim2018} and Fig.~4f therein). In resonators based on Bernal stacked bilayer graphene and few-layer graphene (FLG), irreversible resonant frequency jumps upon changing $V_\mathrm{gate}$ have been ascribed to sudden changes in the stress between the layers  \cite{Kim2020}. In other FLG resonators, RFH have been proposed to result from the sliding of the membrane on its support \cite{Ying2022}. In heterostructures composed of FLG and few-layer hexagonal boron nitride, RFH have been shown to be caused by the growth of bubbles at the interface between the two materials (Ref.~\cite{VarmaSangani2022} and Fig.~2c therein). Overall, these anomalous resonant frequency dispersions are indicative of mechanisms at work in 2-D resonators that are far removed from the properties of conventional membranes and plates \cite{Timoshenko1959}.

\begin{figure*}[t]
\centering
\includegraphics{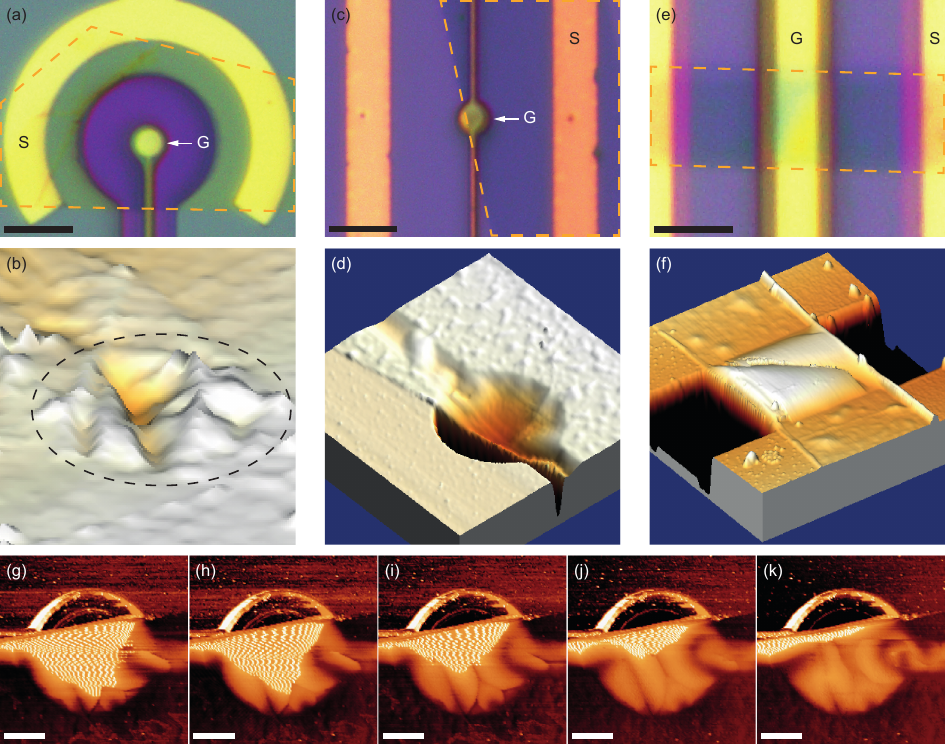}
\caption{Morphologies of crumpled FLG resonators. (a) Optical microscopy (OM) image of Resonator~A. The purple-shaded structure is a mesa in silicon oxide. S (G): source (gate) electrode. Scale bar: 5~$\mu$m. (b) Atomic force microscope (AFM) topography image of Resonator~A (no device annealing). The dashed ellipse identifies the edge of the cavity. (c) OM image of Resonator~B. Scale bar: 5~$\mu$m. (d) AFM image of Resonator~B (after device annealing). (e) OM image of Resonator~C. Scale bar: 3~$\mu$m. (f) AFM image of Resonator~C (after device annealing). (g)-(k) Mapping of the phase of the AFM signal for Resonator~B at $V_\mathrm{gate}=0$~V (g), 2~V (h), 4~V (i), 6~V (j), and 8~V (k). Scale bar: 1~$\mu$m. The phase ranges from $-15^\circ$ (dark) to $+20^\circ$ (bright). In (a), (c), (e), the dashed polygon outlines the FLG flake. In (b), (d), (f), in-plane and out-of plane axes have the same scale (dips and peaks extend $\sim100$~nm above and below the clamping substrate).}\label{Fig1}
\end{figure*}

Here, we report on hystereses in the static response and in the linear dynamic response of four nanomechanical resonators based on crumpled FLG \cite{Liu2011,Kim2011,Zang2013,SDeng2016,LopezPolin2022}. All measurements are performed with the resonators at room temperature and in vacuum. Our measure of the static response is the optical reflectance $R$ of the device composed of the membrane, the gate and the vacuum gap separating the two. Within a plane wave model, $R$ depends on the static, out-of-plane displacement $z_\mathrm{s}$ of the membrane \cite{Chen2018}. Alternatively, we measure the off-resonance background of the spectrum of the driven vibrations, $S_{VV}\propto\vert\partial R/\partial z\vert^2_{z_\mathrm{s}}$, where $z$ is a displacement variable in the out-of-plane direction. We extract the vibrational resonant frequencies $f_\mathrm{n}$ from the peak frequencies of the amplitude spectrum. We report two observations. (i) We find that $R(V_\mathrm{gate})$, $S_{VV}(V_\mathrm{gate})$ and $f_\mathrm{n}(V_\mathrm{gate})$ are hysteretic, with discontinuities in all responses occurring at the same $V_\mathrm{gate}$ values. (ii) We find that the annealing of the devices, either at a mild temperature on a hot plate or \textit{in-situ} via Joule heating, makes the hystereses disappear. We qualitatively discuss our findings and hypothesize that adsorbates play a role in the hysteretic nature of the responses.

\begin{figure*}[t]
\centering
\includegraphics{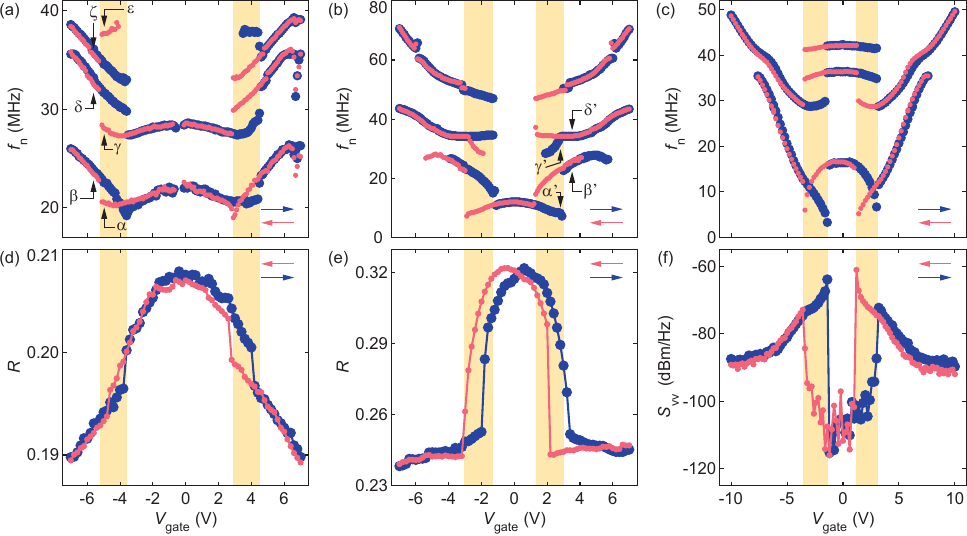}
\caption{Dynamic ($f_\mathrm{n}$) and static ($R$ and $S_{VV}$) responses of crumpled FLG resonators as a function of gate voltage $V_\mathrm{gate}$. Resonant frequencies $f_\mathrm{n}(V_\mathrm{gate})$ upon incrementing (blue dots) and decrementing (pink dots) $V_\mathrm{gate}$ measured in Resonator~A (a), in Resonator~B (b), and in Resonator~C (c). Optical reflectance $R$ upon incrementing (blue dots) and decrementing (pink dots) $V_\mathrm{gate}$ for Resonator~A (d) and for Resonator~B (e). (f) Off-resonance background $S_{VV}(V_\mathrm{gate})$ at $f_\mathrm{d}=1$~MHz of Resonator~C. The yellow-shaded areas in (a-f) highlight the hystereses in $f_\mathrm{n}(V_\mathrm{gate})$. In (a, d), the drive power at the gate is $P_\mathrm{d}=-50$~dBm, and the optical power incident on the resonator is $P_\mathrm{inc}\simeq35$~$\mu$W. In (b, e), $P_\mathrm{d}=-20$~dBm, $P_\mathrm{inc}\simeq31$~$\mu$W. In (c, f), $P_\mathrm{d}=-43$~dBm, $P_\mathrm{inc}\simeq25$~$\mu$W. Greek letters and arrows in (a), (b) indicate the $f_\mathrm{n}(V_\mathrm{gate})$ trace and the value of $V_\mathrm{gate}$ for which mode shapes are measured in Fig.~\ref{Fig2B}.}\label{Fig2}
\end{figure*}

We start by presenting the morphologies of our crumpled membranes transferred onto substrates of different shapes. Figure~\ref{Fig1} shows optical microscopy images and atomic force microscopy (AFM) images of the resonators. Resonator~A consists of a membrane suspended over a cylindrical cavity etched in a raised mesa (Figs.~\ref{Fig1}a, b). The membrane in Resonator~B partially covers a cylindrical cavity in a flat substrate (Figs.~\ref{Fig1}c, d). Resonator~C is made by transferring the FLG flake over a trench (Figs.~\ref{Fig1}e, f). The morphology of a fourth resonator, Resonator~D, is shown in Supplementary Note~1. A local gate electrode is patterned at the bottom of each cavity. The distance between the clamping surface and the gate is nominally 250~nm. All four membranes are crumpled, showing folds, creases and wrinkles extending $\sim100$~nm above and below the clamping surfaces. Figures~\ref{Fig1}g-k display a mapping of the phase of the AFM signal for Resonator~B at several values of $V_\mathrm{gate}$. The fuzzy lines originate from the oscillating AFM cantilever weakly driving the membrane far below resonance \cite{Schwarz2016}; the irregular shape of their distribution indicates that the low frequency part of the mechanical susceptibility $\chi$ of the membrane is nonuniform. At large $V_\mathrm{gate}$, the phase mapping reveals nonuniform patterns within the suspended membrane resembling creases (Fig.~\ref{Fig1}k). We believe that membrane crumpling is caused by fabrication issues, such as shear stress within the polymer stamp during the transfer of the membranes \cite{Liu2011,Kim2011,Zang2013}.

We measure the dynamic and the static responses of the resonators optically \cite{Storch2018,Bunch2007,Davidovikj2016}, as follows. The device is placed in an optical standing wave. As the membrane vibrates, the amount of optical energy it absorbs is modulated. This results in the modulation of the intensity of the light reflected by the device, which is measured with a photodetector. When the resonator is driven by a voltage at frequency $f_\mathrm{d}$ applied between the membrane and the gate, the voltage at the output of the photodetector $\delta V$ oscillates in time $t$ as \cite{Lu2021}:
\begin{equation}
\delta V(t)=P_\mathrm{inc}\left[R\vert_{z_\mathrm{s}}+\frac{\partial R}{\partial z}\Big\vert_{z_\mathrm{s}}\delta z(t)\right]\ast h(t) + \delta u\,,\label{eq1}
\end{equation}
where $P_\mathrm{inc}$ is the optical power incident on the resonator, $\delta z$ is the vibrational amplitude, $h$ is the response of the photodetector, and $\delta u$ is a voltage noise. In a crumpled membrane, $R$ is the average of the spatially-varying reflectance of the device over the cross-section of the laser beam, and $\delta z$ is the averaged amplitude of the vibrations originating from the collective actuation of membrane domains with different mechanical susceptibilities. We measure the output power $P(f_\mathrm{d})=\langle\delta V^2(t)\rangle/(4\times50)$ with a spectrum analyzer. $f_\mathrm{n}$ are the peak frequencies in $P(f_\mathrm{d})$. Figures~\ref{Fig2}a-c show $f_\mathrm{n}(V_\mathrm{gate})$ for Resonators~A, B and C measured upon increasing (blue dots) and decreasing (pink dots) $V_\mathrm{gate}$, from which we make three observations. Firstly, $f_\mathrm{n}(V_\mathrm{gate})$ traces are hysteretic over certain intervals of $V_\mathrm{gate}$ (yellow-shaded areas), at the edges of which $f_\mathrm{n}$ jump. The statistical distributions of the gate voltages $V_\mathrm{gate}^\mathrm{jump}$ at which the jumps occur are shown in Supplementary Note~2. Secondly, these hystereses are nearly symmetrical about $V_\mathrm{gate}\simeq0$, so they depend on the history of the static force $\propto\vert V_\mathrm{gate}\vert$. Thirdly, the frequency $f_0$ of the fundamental modes decreases as $\vert V_\mathrm{gate}\vert$ increases from 0, indicating a softening of the modal spring constant $k_0$; in Fig.~\ref{Fig2}c, $k_0$ nearly dips to zero. As $\vert V_\mathrm{gate}\vert$ increases past $V_\mathrm{gate}^\mathrm{jump}$, $f_0$ increases, signaling a sudden hardening of $k_0$ at the edge of the hysteresis. Interestingly, we observe concomitant hystereses in $R(V_\mathrm{gate})$ measured in Resonators~A and B (Figs.~\ref{Fig2}d, e) and in the off-resonance background $S_{VV}(V_\mathrm{gate})$ of Resonator ~C (Fig.~\ref{Fig2}f). The latter is defined as
\begin{equation}
S_{VV}=P_\mathrm{inc}^2\Big\vert\frac{\partial R}{\partial z}\Big\vert^2_{z_\mathrm{s}}\vert\chi\vert^2F^2\vert H\vert^2+S_{uu}\,,\label{eq2}
\end{equation}
with $F$ the spatially-averaged amplitude of the driving force, $H$ the frequency response of the photodetector, and $S_{uu}$ a noise background. We show $f_\mathrm{n}(V_\mathrm{gate})$ and $R(V_\mathrm{gate})$ for Resonator~D in Supplementary Note~1. 

\begin{figure}[t]
\centering
\includegraphics{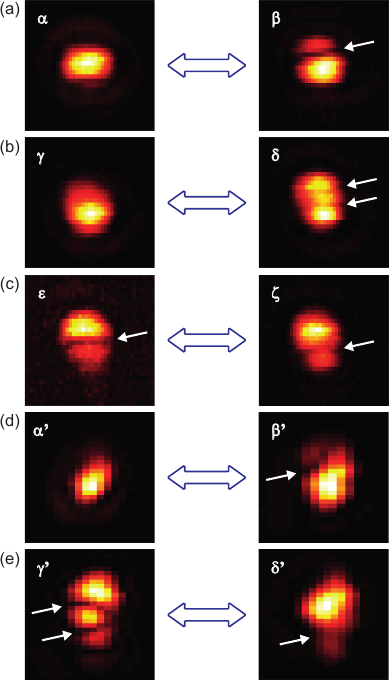}
\caption{Measuring the vibrational mode shapes. (a-c) Mapping of the photodetector output power $P(f_\mathrm{n})$ for Resonator~A at $V_\mathrm{gate}$ indicated by the arrows and the greek letters in Fig.~\ref{Fig2}a, pink dots. Drive power at the gate: $P_\mathrm{d}=-50$~dBm. Optical power incident on the resonator: $P_\mathrm{inc}\simeq35$~$\mu$W. (d, e) Mapping of $P(f_\mathrm{n})$ for Resonator~B at $V_\mathrm{gate}$ indicated by the arrows and the greek letters in Fig.~\ref{Fig2}b, blue dots. $P_\mathrm{d}=-20$~dBm, $P_\mathrm{inc}\simeq31$~$\mu$W. Each map is 4~$\mu$m on a side. White arrows point at vibrational nodes.} \label{Fig2B}
\end{figure}

Furthermore, the jumps in $f_\mathrm{n}$ are accompanied by the onset of new modes. By mapping $P(f_\mathrm{n})$ as a function of the position of the laser beam on the membrane \cite{Davidovikj2016,Lu2021}, we find that the mode shape for a given $f_\mathrm{n}(V_\mathrm{gate})$ trace changes noticeably on opposite sides of a frequency jump [$\alpha\leftrightarrow\beta$, $\gamma\leftrightarrow\delta$, $\epsilon\leftrightarrow\zeta$ in Figs.~\ref{Fig2B}a-c for Resonator A; $\alpha^\prime\leftrightarrow\beta^\prime$, $\gamma^\prime\leftrightarrow\delta^\prime$ in Figs.~\ref{Fig2B}d, e for Resonator B], with the appearance and the disappearance of nodes (white arrows) and with the enlargement of the area of the fundamental modes $\beta$ and $\beta^\prime$. These changes indicate that modes are annihilated and different modes are created across the mechanical transition.

\begin{figure*}[t]
\centering
\includegraphics{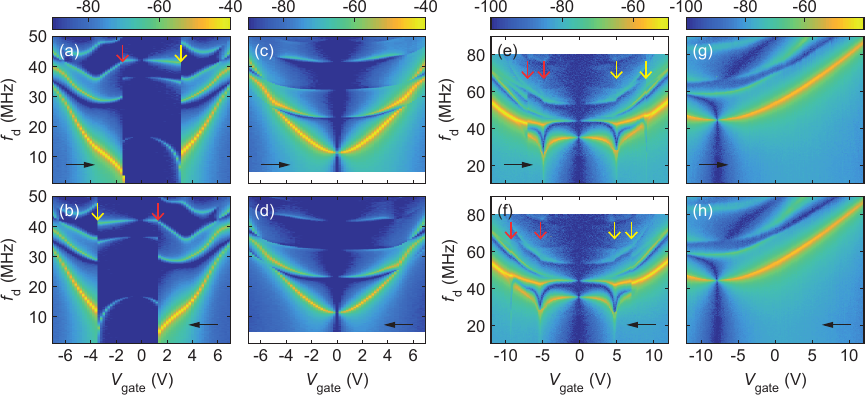}
\caption{Removal of the frequency hystereses by annealing. (a, b) Photodetector output power $P(f_\mathrm{d})$ as a function of drive frequency $f_\mathrm{d}$ for Resonator~C upon increasing (a) and decreasing (b) $V_\mathrm{gate}$ before annealing. (c, d) $P(f_\mathrm{d})$ upon increasing (c) and decreasing (d) $V_\mathrm{gate}$ after hot plate annealing of Resonator~C. In (a-d), the drive power at the gate is $P_\mathrm{d}=-43$~dBm. (e, f) $P(f_\mathrm{d})$ upon increasing (e) and decreasing (f) $V_\mathrm{gate}$ for Resonator~D before annealing. $P_\mathrm{d}=-23$~dBm. (g, h) $P(f_\mathrm{d})$ upon increasing (g) and decreasing (h) $V_\mathrm{gate}$ after current annealing of Resonator~D in the current saturation regime. The minimum of the frequency dispersion is shifted to negative $V_\mathrm{gate}$ due to the electromigration of atoms from the electrodes. $P_\mathrm{d}=-13$~dBm. The vertical arrows in (a, b) and (e, f) point at the transitions. (a-h) Color bars: $P$ (dBm).}\label{Fig4}
\end{figure*}

We find that the annealing of the devices makes the hystereses disappear. After measuring the vibrational spectrum of Resonator~C upon increasing (Fig.~\ref{Fig4}a) and decreasing $V_\mathrm{gate}$ (Fig.~\ref{Fig4}b), we anneal the resonator on a hot plate at 150~$^\circ$C in air for 30~minutes, then we measure its spectrum again (Figs.~\ref{Fig4}c, d). Similarly, we hot-plate anneal Resonator~B and measure its spectrum (Supplementary Note~3). Alternatively and for convenience, we anneal Resonator~D \textit{in-situ} by passing a large current through it \cite{Moser2007}, resulting in the suspended part of the device reaching significantly higher temperatures (Supplementary Note~4). In all three resonators, we observe that $f_\mathrm{n}(V_\mathrm{gate})$ and $S_{VV}(V_\mathrm{gate})$ are no longer hysteretic after annealing. Furthermore, the frequency dispersions at low $V_\mathrm{gate}$ have been strongly modified. Namely, spring constant softening before annealing has been replaced with spring constant hardening after annealing. By contrast, the behavior of $f_\mathrm{n}(V_\mathrm{gate})$ for $\vert V_\mathrm{gate}\vert>\vert V_\mathrm{gate}^\mathrm{jump}\vert$ has remained nearly unchanged.

Our observations bring to mind some of the behaviors reported in Ref.~\cite{VarmaSangani2022}, where resonators based on a layer of hexagonal boron nitride stacked on graphene are measured. As in our work, hystereses in $f_\mathrm{n}(V_\mathrm{gate})$ are found (see Fig.~2c in Ref.~\cite{VarmaSangani2022}), and the appearance of new modes is observed across a transition (see Fig.~4f in Ref.~\cite{VarmaSangani2022}). We find the behavior of Device D3 in Ref.~\cite{VarmaSangani2022} to be especially interesting: there, the dispersion of $f_0$ is negative at lower $V_\mathrm{gate}$ and becomes positive after a transition at higher $V_\mathrm{gate}$. This behavior is interpreted as the delamination of the heterostructure. At low $V_\mathrm{gate}$, the spring constant is dominated by capacitive softening because the tensile stress is high. As $\vert V_\mathrm{gate}\vert$ increases, the two membranes are pulled apart, lowering the interfacial stress, until the transition is reached where the two low-stress membranes become free to vibrate independently. Negative dispersions followed by a revival of positive dispersions across the transition are also observed in our crumpled FLG resonators. 

While the complicated membrane morphologies prevent us from offering a theoretical model to interpret our findings quantitatively (Supplementary Note~5), hystereses in both the static and the dynamic responses, along with the fact that all $f_\mathrm{n}(V_\mathrm{gate})$ traces show a discontinuity at the same $V_\mathrm{gate}^\mathrm{jump}$, indicate that the shape of the membranes and their tensile strain undergo hysteretic changes. Because the hystereses are sharp ($f_\mathrm{n}$, $R$ and $S_{VV}$ jump across the transition), these changes are unlikely to originate from the sliding of the membranes on their substrate \cite{Ying2022}. We also rule out slips between layers, as observed in FLG resonators \cite{Kim2020}, since those slips have been shown to yield jumps in $f_\mathrm{n}$ upon changing $V_\mathrm{gate}$ that are not recoverable when $V_\mathrm{gate}$ is cycled back. Alternatively, given that a mild temperature annealing removes the hystereses, a plausible explanation is that physisorbed molecules, such as water \cite{Brandenburg2019}, play a role in the hysteretic responses. Because water is known to promote the adhesion of graphene \cite{Ma2017}, we hypothesize that water molecules make the surfaces of a folded membrane stick together, enhancing the built-in strain $\epsilon_\textrm{built-in}$ within the membrane and making it the main contribution to the total strain, $\epsilon\simeq\epsilon_\textrm{built-in}$. In turn, the dependence on $V_\mathrm{gate}$ of the spring constant $k_0$ of the fundamental mode of a graphene membrane under tensile stress is dominated by capacitive softening \cite{Weber2014}, $k_0(V_\mathrm{gate})=k_{0,\mathrm{el}}+k_\mathrm{cap}(V_\mathrm{gate})$. Here, $k_{0,\mathrm{el}}$ is the elastic spring constant of the fundamental mode, which depends on $V_\mathrm{gate}$ only weakly, and $k_\mathrm{cap}\propto-V_\mathrm{gate}^2\vert\partial^2C_\mathrm{gate}/\partial z^2\vert_{z_\mathrm{s}}$; $C_\mathrm{gate}$ is the capacitance between the membrane and the gate, which depends on the static shape of the membrane \cite{Lifshitz2008}. Such a spring softening behavior is what we qualitatively observe in $f_0(V_\mathrm{gate})=1/(2\pi)[k_0(V_\mathrm{gate})/m]^{1/2}$, with $m$ the mass of the fundamental mode, before annealing and for $\vert V_\mathrm{gate}\vert<\vert V_\mathrm{gate}^\mathrm{jump}\vert$. An increasing static force $\propto\vert V_\mathrm{gate}\vert$ may pull the folds open, adding some vibrational area, as we observe. The opening of the folds would reduce $\epsilon_\textrm{built-in}$ to a low enough value $\tilde{\epsilon}_\textrm{built-in}\ll\epsilon_\textrm{built-in}$ that the dependence of $\epsilon$ on $V_\mathrm{gate}$ may become sizeable, $\epsilon=\tilde{\epsilon}_\textrm{built-in}+\tilde{\epsilon}(V_\mathrm{gate})$. This would endow the elastic energy of the membrane $U_\mathrm{el}(\epsilon)$ with its own $V_\mathrm{gate}$ dependence. In turn, $k_{0,\mathrm{el}}=\partial^2U_\mathrm{el}/\partial z^2\vert_{z_\mathrm{s}}$ would increase with $\vert V_\mathrm{gate}\vert$, because $U_\mathrm{el}$ is mostly a stretching energy that increases with $z_\mathrm{s}(V_\mathrm{gate})$ (as explained in Refs.~\cite{Storch2018,Chen2009,Barton2012}).

The hysteretic behavior of $k_0(V_\mathrm{gate})$ upon cycling $\vert V_\mathrm{gate}\vert$ may indicate a gradual ``unsticking'' of folded graphene surfaces that is complete at a larger $\vert V_\mathrm{gate}\vert$, followed by a gradual ``resticking'' that is achieved at a lower $\vert V_\mathrm{gate}\vert$. After annealing, we hypothesize that the folds remain open. The fact that little changes in $f_\mathrm{n}(V_\mathrm{gate})$ are observed before and after annealing for $\vert V_\mathrm{gate}\vert>\vert V_\mathrm{gate}^\mathrm{jump}\vert$, where folds are open in both cases, supports the idea that the annealing simply removes adsorbates acting as an adhesive. The opening and closing of folds in our mechanism resemble the partial delamination of the heterostructures in Ref.~\cite{VarmaSangani2022}. Our contribution is to qualitatively address the role of adsorbates in this mechanism and to highlight the effect of the annealing on the responses.

Much work remains to quantify the effect of adsorbates trapped in a crumpled graphene membrane on the mechanical properties of the membrane. This includes measuring the dependence of $k_0$ on static displacement, which is expected to be nonlinear \cite{RuizVargas2011,Nelson2013,Nicholl2015,Gornyi2017,Nicholl2017}. This also entails measuring the Young's modulus $E$ of the membrane, given that $E$ has been shown to be significantly smaller in crumpled FLG \cite{Nicholl2015,RuizVargas2011,Nicholl2017} than in uncrumpled FLG \cite{Frank2007,Lee2008}. Exposing crumpled FLG to moisture and to various organic gases, post annealing, and verifying that RFH can be recovered, would provide a form of detection of chemical species \cite{Kang2016}. Measuring charges and dipoles \cite{Chin2021} through electrostatic force microscopy \cite{Moser2008} may also advance our understanding of crumpled resonators. Finally, locally annealing folds with a tightly focused laser beam in larger resonators and at zero $V_\mathrm{gate}$, then measuring changes in the vibrational spectra, may provide an estimate of their interfacial energy.

See the supplementary material for the hystereses in the responses of Resonator~D, the distribution of gate voltages at which resonant frequencies jump, the effect of annealing on the dynamic response of Resonator~B, the current annealing of Resonator~D, and an attempt at reproducing the anomalous frequency dispersions with a regular membrane model.

\section*{Acknowledgement}

This work was supported by the National Natural Science Foundation of China (grant numbers 62074107 and 62150710547) and the project of the Priority Academic Program Development (PAPD) of Jiangsu Higher Education Institutions. J. M. acknowledges inspiring discussions with A. Reserbat-Plantey and F. Vialla. The authors are grateful to Prof. Wang Chinhua for his strong support.

\section*{Supplementary Material}

\subsection{Hystereses in the responses of Resonator~D}

\begin{figure*}[t] 
\centering
\includegraphics[width=12cm]{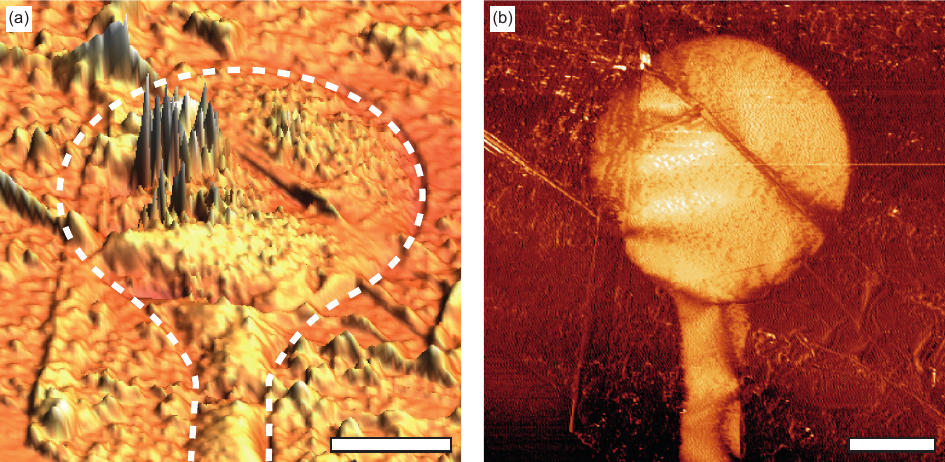}
\caption{Atomic force microscopy (AFM) images of Resonator~D. (a) Topography image. Except for the sharp peaks in the top left corner, which are an experimental artefact (see main text), the maximum height of the membrane corrugation is 15~nm. Scale bar: 1~$\mu$m. The white dashed line follows the contour of the cavity. (b) Mapping of the phase of the AFM signal. Scale bar: 1~$\mu$m. The phase ranges from $-10^\circ$ (dark) to $+15^\circ$ (bright).}\label{FigS1afm}
\end{figure*}

Figure~\ref{FigS1afm} displays atomic force microscopy (AFM) images of Resonator~D. Both the topography image (Fig.~\ref{FigS1afm}a) and a mapping of the phase of the AFM signal (Fig.~\ref{FigS1afm}b) are shown, revealing that the membrane is crumpled.

\begin{figure}[H]
\centering
\includegraphics[width=70mm]{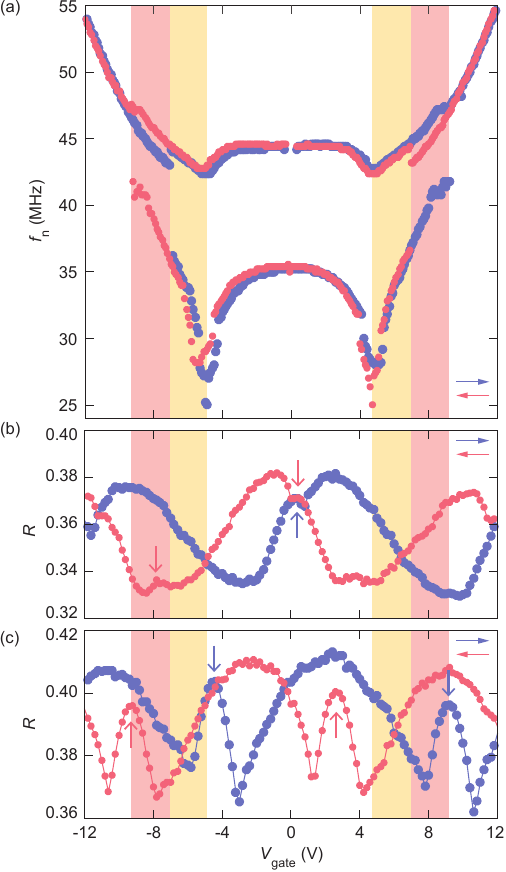}
\caption{Hystereses in the dynamic response and in the static response of Resonator~D. (a) Dispersion of the vibrational resonant frequencies $f_\mathrm{n}(V_\mathrm{gate})$. (b) Optical reflectance $R(V_\mathrm{gate})$ [first dataset]. (c) $R(V_\mathrm{gate})$ [second dataset]. Drive power applied to the gate: $P_\mathrm{d}=-23$~dBm. Optical power incident on the resonator: $P_\mathrm{inc}\simeq35$~$\mu$W. In (b, c), vertical arrows point at finer features. The two differently shaded areas highlight two hystereses.}\label{FigS1}
\end{figure}

The hysteretic responses of Resonator~D are shown in Fig.~\ref{FigS1}. Figure~\ref{FigS1}a shows the dispersion of the vibrational resonant frequencies $f_\mathrm{n}$ as a function of increasing and decreasing gate voltage $V_\mathrm{gate}$. The data show several $V_\mathrm{gate}$ intervals within which $f_\mathrm{n}$ are hysteretic. The two most visible hystereses are highlighted by the shaded areas. Figures~\ref{FigS1}b, c show two successive measurements of the optical reflectance $R$ as a function of $V_\mathrm{gate}$. The two datasets reveal a similar trend, with finer features barely emerging in Fig.~\ref{FigS1}b and appearing as additional oscillations in Fig.~\ref{FigS1}c [see vertical arrows in panels (b, c)]. The emergence of these finer features may indicate gradual changes in the shape of the membrane.

\subsection{Distribution of gate voltages at which resonant frequencies jump}

\begin{figure}[t]
\centering
\includegraphics{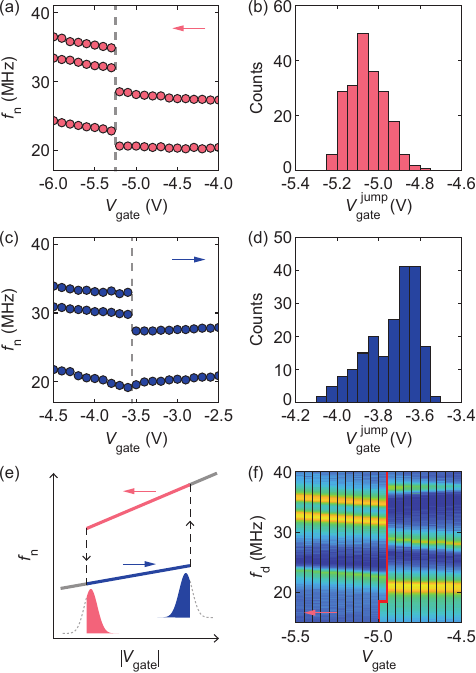}
\caption{Statistics of frequency jumps for Resonator~A. (a) Single instance of $f_\mathrm{n}(V_\mathrm{gate})$ and (b) distribution of $V_\mathrm{gate}^\mathrm{jump}$ at which values $f_\mathrm{n}$ jump upon decreasing $V_\mathrm{gate}$. (c) Single instance of $f_\mathrm{n}(V_\mathrm{gate})$ and (d) distribution of $V_\mathrm{gate}^\mathrm{jump}$ upon increasing $V_\mathrm{gate}$. In (a) and (c), the vertical dashed line indicates the transition. (e) Illustration of the ``latching switch effect'' of the hysteresis, which results in the distributions of $V_\mathrm{gate}^\mathrm{jump}$ being skewed. (f) Measured vibrational power spectra (black boxes) where vibrations are driven by a signal at frequency $f_\mathrm{d}$ and power $P_\mathrm{d}=-50$~dBm. $V_\mathrm{gate}$ is stepped by 50~mV towards $-5.5$~V (pink arrow). The thick red line indicates the transition evidenced by a sudden change in background power. Here, $V_\mathrm{gate}^\mathrm{jump}=-4.975$~V. Blue (yellow): $P\simeq-85$~dBm ($P\simeq-55$~dBm).} \label{Fig_statistics}
\end{figure}

Jumps in the resonant frequencies $f_\mathrm{n}$ statistically occur within an asymmetric distribution of gate voltages $V_\mathrm{gate}^\mathrm{jump}$ whose skewness depends on the direction along which $V_\mathrm{gate}$ is changed. Figure~\ref{Fig_statistics}a shows one instance of $f_\mathrm{n}(V_\mathrm{gate})$ and Fig.~\ref{Fig_statistics}b shows the distribution of $V_\mathrm{gate}^\mathrm{jump}$ where $V_\mathrm{gate}$ is decremented (Resonator~A). Figures~\ref{Fig_statistics}c, d show the corresponding measurements obtained upon incrementing $V_\mathrm{gate}$. The large width of the distributions, $W\simeq400$~mV, indicates the presence of fluctuations within the resonator. These may originate from stochastic reconfigurations of mesoscopic few-layer graphene domains, which are connected by folds acting as hinges and springs \cite{LopezPolin2022}. Further, the fact that all $f_\mathrm{n}(V_\mathrm{gate})$ traces show a discontinuity at a given $V_\mathrm{gate}^\mathrm{jump}$ indicates that a transition affecting the whole membrane has occurred.

We propose that the distribution of $V_\mathrm{gate}^\mathrm{jump}$ is shaped by the ``latching switch effect'' of the hysteretic response at the transition. We illustrate this statement in Fig.~\ref{Fig_statistics}e where, for simplicity, mechanical fluctuations have been converted into effective $V_\mathrm{gate}$ fluctuations. As $V_\mathrm{gate}$ approaches the transition, the tail of the distribution starts triggering rare frequency jumps, which become increasingly likely as the maximum of the distribution nearly matches the transition. Beyond that point, the distribution of $V_\mathrm{gate}^\mathrm{jump}$ steeply decreases as the transition has already occurred and the hysteresis precludes fluctuations from returning the membrane to its earlier state. This ``latching switch effect'' is visible in the spectrum measurements where $f_\mathrm{d}$ is swept: only one jump in $f_\mathrm{n}$ is observed at the transition, occurring some time after $V_\mathrm{gate}$ has been set (thick red line in Fig.~\ref{Fig_statistics}f). In effect, the hysteresis acts as a threshold detector for fluctuations near the transition, which are otherwise not seen (e.g. as slow frequency noise) in the spectral resonances.

\subsection{Effect of annealing on the dynamic response of Resonator~B}

\begin{figure}[t]
\centering
\includegraphics{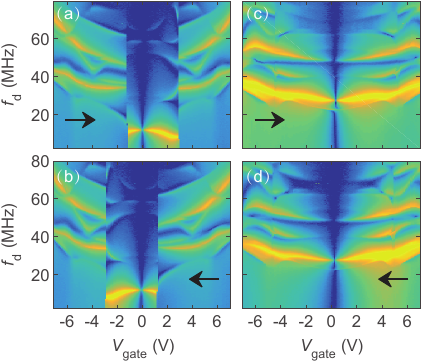}
\caption{Removal of the hystereses in the dynamic response of Resonator~B by hot plate annealing. Vibrational power spectra upon increasing (a) and decreasing (b) the gate voltage $V_\mathrm{gate}$ prior to annealing. Power spectra upon increasing (c) and decreasing (d) $V_\mathrm{gate}$ after annealing. Drive power at the gate $P_\mathrm{d}=-20$~dBm. Optical power incident on the resonator $P_\mathrm{inc}\simeq31$~$\mu$W. Vibrational power scale: Blue, $P=-90$~dBm; Yellow, $P=-40$~dBm.}\label{FigS2}
\end{figure}

Figure~\ref{FigS2} shows the effect of hot plate annealing on the vibrational spectrum of Resonator~B. The spectrum is the power $P$ of the signal at the output of the photodetector measured as a function of drive frequency $f_\mathrm{d}$ over a range of increasing (a, c) and decreasing (b, d) gate voltage $V_\mathrm{gate}$. Spectra in (c) and (d) are measured after the annealing of the device in air on a hot plate at 150$^\circ$C for 30~minutes. While frequency hystereses are visible in (a, b) as $V_\mathrm{gate}$ is increased and then decreased, the responses post annealing are non-hysteretic.

\subsection{Current annealing of Resonator~D}

\begin{figure*}[t]
\centering
\includegraphics{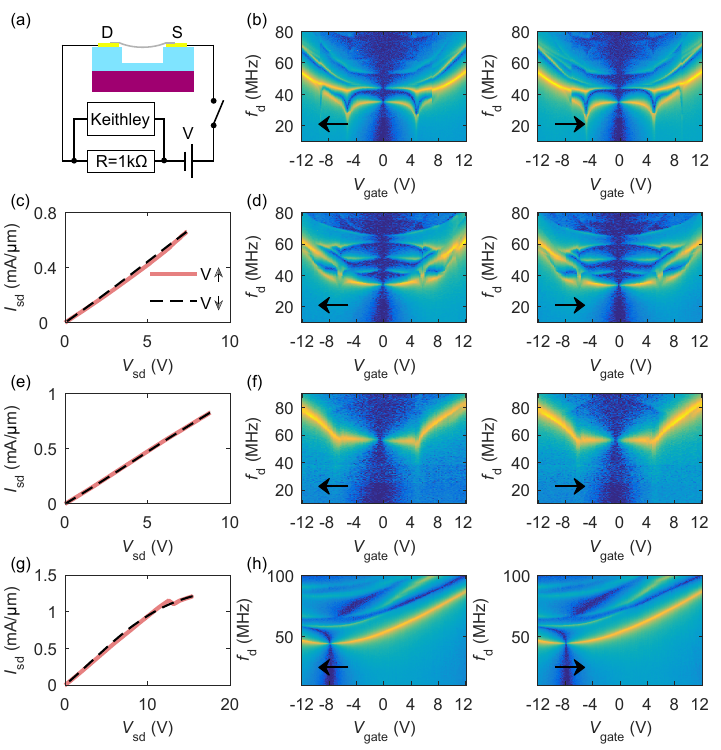}
\caption{Removal of the hystereses in the dynamic response by current annealing of Resonator~D. (a) Schematic of the current annealing setup. ``Keithley'' is a high input impedance voltmeter. $V$ is a dc voltage applied across the equivalent resistor composed of a resistor $R=1$~kOhm in series with the resistor at the interface between the source electrode S and the resonator, the resonator, and the resistor at the interface between the resonator and the drain electrode D. (b) Vibrational power spectra upon decreasing (left) and increasing (right) gate voltage $V_\mathrm{gate}$ prior to the annealing of the device. (c) Current density $I_\mathrm{sd}$ (in units of mA per width of the membrane along the direction perpendicular to the current flow) in the circuit as a function of voltage drop across source and drain $V_\mathrm{sd}$ upon increasing and decreasing $V$. (d) Power spectra measured after this first annealing. (e) $I_\mathrm{sd}(V_\mathrm{sd})$ for the second current annealing. (f) Power spectra measured after the second annealing. (g) $I_\mathrm{sd}(V_\mathrm{sd})$ for the third current annealing. (h) Power spectra measured after the third annealing.}\label{FigS3}
\end{figure*}

Figure~\ref{FigS3} shows the effect of the current annealing of Resonator~D on the vibrational spectrum of the resonator. The current annealing setup is shown in Fig.~\ref{FigS3}a. The spectrum prior to annealing is shown in Fig.~\ref{FigS3}b. The first current annealing is shown in Fig.~\ref{FigS3}c and the resulting spectrum is shown in Fig.~\ref{FigS3}d. A second annealing is depicted in Figs.~\ref{FigS3}e, f, and a third annealing is shown in Figs.~\ref{FigS3}g, h. The minimum of the frequency dispersion in Fig.~\ref{FigS3}h is shifted to negative $V_\mathrm{gate}$, presumably as a result of the electromigration of atoms from the electrodes onto the resonator.

In the current saturation regime, the central area of the resonator is expected to reach high temperatures. In resonators with a similar geometry as ours, and using similar source-drain currents and source-drain voltages, the authors of Refs.~\cite{Dorgan2013,Ye2018} measure temperatures exceeding 1000~K.

\subsection{Attempt at reproducing the anomalous frequency dispersions with a regular membrane model}

\begin{figure}[t]
\centering
\includegraphics{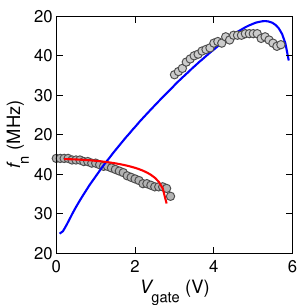}
\caption{Attempt at reproducing the anomalous frequency dispersions of Resonator~B with a regular membrane model. We consider the fundamental mode, $f_\mathrm{n}=f_0$.}\label{FigS4}
\end{figure}

Here, we show that reproducing the frequency dispersions displayed by Resonator~B in the main text with a regular membrane model \cite{Storch2018,Chen2018} would require adjusting the membrane's built-in strain $\epsilon_\textrm{built-in}$ across the transitions (vibrational frequency jumps at certain gate voltages) and employing unrealistically low two-dimensional Young's modulii $E$. Figure~\ref{FigS4} displays the measured frequency dispersion upon increasing the gate voltage $V_\mathrm{gate}$ (dots), along with two calculated frequency dispersions. In both calculations, we consider a membrane made of 9 graphene layers (estimated from optical reflectometry of the membrane in contact with electrodes \cite{Lu2021}), a cavity radius of 1.55~$\mu$m, and a nominal cavity depth of 170~nm (the smallest distance between the membrane and the top surface of the gate measured by AFM). In the model, the membrane is flat at zero $V_\mathrm{gate}$, which is clearly not the case in the experiment. We calculate the resonant frequency of the fundamental vibration mode $f_0$ by minimizing the total energy (the sum of the stretching energy of the membrane and the electrical energy stored in the capacitor formed by the membrane and the gate electrode) and by computing the second derivative of the total energy with respect to membrane dispacement. For the dispersion in red (lower $V_\mathrm{gate}$), we use $\epsilon_\textrm{built-in}=3\times10^{-2}$ and $E=2\times10^8$~Pa. For the dispersion in blue (higher $V_\mathrm{gate}$), we use $\epsilon_\textrm{built-in}=4\times10^{-5}$ and $E=7\times10^9$~Pa. The caveat to this analysis, however, is that (i) $\epsilon_\textrm{built-in}=3\times10^{-2}$ is rather large (a maximum of $\epsilon_\textrm{built-in}=6\times10^{-2}$ was found for free-standing, single-crystalline monolayer graphene grown by chemical vapor deposition \cite{Cao2020,Jaddi2024}), and (ii) $E=7\times10^9$~Pa is 10 times smaller than the smallest $E$ measured in crumpled single layer graphene at room temperature \cite{Nicholl2015}.

\begin{thebibliography}{00}
\bibitem{Storch2018}
I. R. Storch, R. De Alba, V. P. Adiga, T. S. Abhilash, R. A. Barton, H. G. Craighead, J. M. Parpia, and P. L. McEuen, "Young’s modulus and thermal expansion of tensioned graphene membranes," \textit{Phys. Rev. B} \textbf{98}(8), 085408 (2018). \url{https://doi.org/10.1103/PhysRevB.98.085408}

\bibitem{Sazonova2004}
V. Sazonova, Y. Yaish, H. \"{U}st\"{u}nel, D. Roundy, T. A. Arias, and P. L. McEuen, "A tunable carbon nanotube electromechanical oscillator," \textit{Nature} \textbf{431}(7006), 284--287 (2004). \url{https://doi.org/10.1038/nature02905}

\bibitem{Bunch2007}
J. S. Bunch, A. M. van der Zande, S. S. Verbridge, I. W. Frank, D. M. Tanenbaum, J. M. Parpia, H. G. Craighead, and P. L. McEuen, "Electromechanical resonators from graphene sheets," \textit{Science} \textbf{315}(5811), 490--493 (2007). \url{https://doi.org/10.1126/science.1136836}

\bibitem{Chen2009}
C. Chen, S. Rosenblatt, K. I. Bolotin, W. Kalb, P. Kim, I. Kymissis, H. L. Stormer, T. F. Heinz, and J. Hone, "Performance of monolayer graphene nanomechanical resonators with electrical readout," \textit{Nat. Nanotechnol.} \textbf{4}(12), 861--867 (2009). \url{https://doi.org/10.1038/nnano.2009.267}

\bibitem{Ye2021}
F. Ye, A. Islam, T. Zhang, and P. X.-L. Feng, "Ultrawide Frequency Tuning of Atomic Layer van der Waals Heterostructure Electromechanical Resonators," \textit{Nano Lett.} \textbf{21}(13), 5508–5515 (2021). \url{https://doi.org/10.1021/acs.nanolett.1c00610}

\bibitem{Xu2022}
B. Xu, P. Zhang, J. Zhu, Z. Liu, A. Eichler, X.-Q. Zheng, J. Lee, A. Dash, S. More, S. Wu, Y. Wang, H. Jia, A. Naik, A. Bachtold, R. Yang, P. X.-L. Feng, and Z. Wang, "Nanomechanical Resonators: Toward Atomic Scale," \textit{ACS Nano} \textbf{16}(10), 15545--15585 (2022). \url{https://doi.org/10.1021/acsnano.2c01673}

\bibitem{Carr1997}
D. W. Carr and H. G. Craighead, "Fabrication of nanoelectromechanical systems in single crystal silicon using silicon on insulator substrates and electron beam lithography," \textit{J. Vac. Sci. Technol. B} \textbf{15}, 2760-–2763 (1997). \url{https://doi.org/10.1116/1.589722}

\bibitem{Lifshitz2008}
R. Lifshitz and M. C. Cross, "Nonlinear dynamics of nanomechanical and micromechanical resonators," in \textit{Reviews of Nonlinear Dynamics and Complexity}, Vol. 1. (ed. H.G. Schuster) 1--52 (Wiley-VCH, 2008). \url{https://doi.org/10.1002/9783527626359.ch1}

\bibitem{Nicholl2015}
R. J. T. Nicholl, H. J. Conley, N. V. Lavrik, I. Vlassiouk, Y. S. Puzyrev, V. P. Sreenivas, S. T. Pantelides, and K. I. Bolotin, "The effect of intrinsic crumpling on the mechanics of free-standing graphene," \textit{Nat. Commun.} \textbf{6}, 8789 (2015). \url{https://doi.org/10.1038/ncomms9789}

\bibitem{Song2012}
X. Song, M. Oksanen, M. A. Sillanp\"{a}\"{a}, H. G. Craighead, J. M. Parpia, and P. J. Hakonen, "Stamp transferred suspended graphene mechanical resonators for radio frequency electrical readout,"" \textit{Nano Lett.} \textbf{12}(1), 198--202 (2012). \url{https://doi.org/10.1021/nl203305q}

\bibitem{Barton2012}
R. A. Barton, I. R. Storch, V. P. Adiga, R. Sakakibara, B. R. Cipriany, B. Ilic, S.-P. Wang, P. Ong, P. L. McEuen, J. M. Parpia, and H. G. Craighead, "Photothermal self-oscillation and laser cooling of graphene optomechanical systems," \textit{Nano Lett.} \textbf{12}(9), 4681--4686 (2012). \url{https://doi.org/10.1021/nl302036x}

\bibitem{Chen2018}
F. Chen, C. Yang, W. Mao, H. Lu, K. G Schaedler, A. Reserbat-Plantey, J. Osmond, G. Cao, X. Li, C. Wang, Y. Yan, and J. Moser, "Vibration detection schemes based on absorbance tuning in monolayer molybdenum disulfide mechanical resonators," \textit{2D Materials} \textbf{6}(1), 011003 (2018). \url{https://doi.org/10.1088/2053-1583/aae5b7}

\bibitem{Guettinger2017}
J. G\"{u}ttinger, A. Noury, P. Weber, A. M. Eriksson, C. Lagoin, J. Moser, C. Eichler, A. Wallraff, A. Isacsson, and A. Bachtold, "Energy-dependent path of dissipation in nanomechanical resonators," \textit{Nat. Nanotechnol.} \textbf{12}(7), 631-–636 (2017). \url{https://doi.org/10.1038/NNANO.2017.86}

\bibitem{Mathew2016}
J. P. Mathew, R. N. Patel, A. Borah, A., R. Vijay, and M. M. Deshmukh, "Dynamical strong coupling and parametric amplification of mechanical modes of graphene drums," \textit{Nat. Nanotechnol.} \textbf{11}(9), 747--751 (2016). \url{https://doi.org/10.1038/nnano.2016.94}

\bibitem{Mahboob2008}
I. Mahboob and H. Yamaguchi, "Bit storage and bit flip operations in an electromechanical oscillator," \textit{Nat. Nanotechnol.} \textbf{3}(5), 275--279 (2008). \url{https://doi.org/10.1038/nnano.2008.84}

\bibitem{Eichler2011}
A. Eichler, J. Chaste, J. Moser, and A. Bachtold, "Parametric amplification and self-oscillation in a carbon nanotube resonator," \textit{Nano Lett.} \textbf{11}(7), 2699--2703 (2011). \url{https://doi.org/10.1021/nl200950d}

\bibitem{Su2021}
Z.-J. Su, Y. Ying, X.-X. Song, Z.-Z. Zhang, Q.-H. Zhang, G. Cao, H.-O. Li, G.-C. Guo, and G.-P. Guo, "Tunable parametric amplification of a graphene nanomechanical resonator in the nonlinear regime," \textit{Nanotechnology} \textbf{32}(15), 155203 (2021). \url{https://dx.doi.org/10.1088/1361-6528/abc9ea}

\bibitem{Weber2016}
P. Weber, J. G\"{u}ttinger, A. Noury, J. Vergara-Cruz, and A. Bachtold, "Force sensitivity of multilayer graphene optomechanical devices," \textit{Nat. Commun.} \textbf{7}, 12496 (2016). \url{https://doi.org/10.1038/ncomms12496}

\bibitem{Chen2013}
C. Chen, S. Lee, V. V. Deshpande, G.-H. Lee, M. Lekas, K. Shepard, and J. Hone, "Graphene mechanical oscillators with tunable frequency," \textit{Nat. Nanotechnol.} \textbf{8}(12), 923--927 (2013). \url{https://doi.org/10.1038/nnano.2013.232}

\bibitem{Chaste2020}
J. Chaste, I. Hnid, L. Khalil, C. Si, A. Durnez, X. Lafosse, M.-Q. Zhao, A. T. C. Johnson, S. Zhang, J. Bang, and A. Ouerghi, "Phase transition in a memristive suspended MoS$_{2}$ monolayer probed by opto- and electro-mechanics," \textit{ACS Nano} \textbf{14}(10), 13611--13618 (2020). \url{https://doi.org/10.1021/acsnano.0c05721}

\bibitem{Jiang2020}
S. Jiang, H. Xie, J. Shan, and K. F. Mak, "Exchange magnetostriction in two-dimensional antiferromagnets," \textit{Nat. Mater.} \textbf{19}(12), 1295--1299 (2020). \url{https://doi.org/10.1038/s41563-020-0712-x}

\bibitem{Sahu2022}
S. K. Sahu, S. Mandal, S. Ghosh, M. M. Deshmukh, and V. Singh, "Superconducting vortex-charge measurement using cavity electromechanics," \textit{Nano Lett.} \textbf{22}(4), 1665--1671 (2022). \url{https://doi.org/10.1021/acs.nanolett.1c04688}

\bibitem{Kim2018}
S. P. Kim, J. Yu, and A. M. van der Zande, "Nano-electromechanical Drumhead Resonators from Two-Dimensional Material Bimorphs," \textit{Nano Lett.} \textbf{18}(11), 6686--6695 (2018). \url{https://doi.org/10.1021/acs.nanolett.8b01926}

\bibitem{Kim2020}
S. P. Kim, E. Annevelink, E. Han, J. Yu, P. Y. Huang, E. Ertekin, and A. M. van der Zande, "Stochastic Stress Jumps Due to Soliton Dynamics in Two-Dimensional van der Waals Interfaces," \textit{Nano Lett.} \textbf{20}(2), 1201--1207 (2020). \url{https://doi.org/10.1021/acs.nanolett.9b04619}

\bibitem{Ying2022}
Y. Ying, Z.-Z. Zhang, J. Moser, Z.-J. Su, X.-X. Song, and G.-P. Guo, "Sliding nanomechanical resonators," \textit{Nat. Commun.} \textbf{13}, 6392 (2022). \url{https://doi.org/10.1038/s41467-022-34144-5}

\bibitem{VarmaSangani2022}
L. D. Varma Sangani, S. Mandal, S. Ghosh, K. Watanabe, T. Taniguchi, and M. M. Deshmukh, "Dynamics of Interfacial Bubble Controls Adhesion Mechanics in Van der Waals Heterostructure," \textit{Nano Lett.} \textbf{22}(9), 3612--3619 (2022). \url{https://doi.org/10.1021/acs.nanolett.1c04341}

\bibitem{Timoshenko1959}
S. Timoshenko and S. Woinowsky-Krieger, "Theory of Plates and Shells." McGraw-Hill College. 2nd Edition (1959).

\bibitem{Liu2011}
N. Liu, Z. H. Pan, L. Fu, C. H. Zhang, B. Dai, and Z. F. Liu, ``The Origin of Wrinkles on Transferred Graphene,'' \textit{Nano Res.} \textbf{4}(10), 996--1004 (2011). \url{https://doi.org/10.1007/s12274-011-0156-3}

\bibitem{Kim2011}
K. Kim, Z. Lee, B. D. Malone, K. T. Chan, B. Alem\'{a}n, W. Regan, W. Gannett, M. F. Crommie, M. L. Cohen, and A. Zettl, ``Multiply folded graphene,'' \textit{Phys. Rev. B} \textbf{83}(24), 245433 (2011). \url{https://dx.doi.org/10.1103/PhysRevB.83.245433}

\bibitem{Zang2013}
J. Zang, S. Ryu, N. Pugno, Q. Wang, Q. Tu, M. J. Buehler, and X. Zhao, ``Multifunctionality and control of the crumpling and unfolding of large-area graphene,'' \textit{Nature Mater.} \textbf{12}, 321--325 (2013). \url{https://doi.org/10.1038/nmat3542}

\bibitem{SDeng2016}
S. Deng and V. Berry, ``Wrinkled, rippled and crumpled graphene: an overview of formation mechanism, electronic properties, and applications,'' \textit{Mater. Today} \textbf{19}(4), 197--212 (2016). \url{https://dx.doi.org/10.1016/j.mattod.2015.10.002}

\bibitem{LopezPolin2022}
G. Lopez-Polin, C. Gomez-Navarro, and J. Gomez-Herrero, ``The effect of rippling on the mechanical properties of graphene,'' \textit{Nano Materials Science} \textbf{4}, 18--26 (2022). \url{https://doi.org/10.1016/j.nanoms.2021.05.005}

\bibitem{Schwarz2016}
C. Schwarz, "Optomechanical, Vibrational and Thermal Properties of Suspended Graphene Membranes," \textit{Ph. D. Thesis}, Universit\'{e} de Grenoble (2016). \url{https://theses.hal.science/tel-01493121}

\bibitem{Davidovikj2016}
D. Davidovikj, J. J. Slim, S. J. Cartamil-Bueno, H. S. J. van der Zant, P. G. Steeneken, and W. J. Venstra, "Visualizing the motion of graphene nanodrums," \textit{Nano Lett.} \textbf{16}(4), 2768--2773 (2016). \url{https://doi.org/10.1021/acs.nanolett.6b00477}

\bibitem{Lu2021}
H. Lu, C. Yang, Y. Tian, J. Lu, F. Xu, C. Zhang, F. Chen, Y. Yan, K. G. Schaedler, C. Wang, F. H. L. Koppens, A. Reserbat-Plantey, and J. Moser, "Imaging vibrations of electromechanical few layer graphene resonators with a moving vacuum enclosure," \textit{Precis. Eng.} \textbf{72}, 769--776 (2021). \url{https://doi.org/10.1016/j.precisioneng.2021.06.012}

\bibitem{Moser2007}
J. Moser, A. Barreiro, and A. Bachtold, ``Current-induced cleaning of graphene,'' \textit{Appl. Phys. Lett.} \textbf{91}(16), 163513 (2007). \url{https://dx.doi.org/10.1063/1.2789673}

\bibitem{Brandenburg2019}
J. G. Brandenburg, A. Zen, M. Fitzner, B. Ramberger, G. Kresse, T. Tsatsoulis, A. Gr\"{u}neis, A. Michaelides, and Dario Alf\`{e}, ``Physisorption of Water on Graphene: Subchemical Accuracy from
Many-Body Electronic Structure Methods,'' \textit{J. Phys. Chem. Lett.} \textbf{10}, 358--368 (2019). \url{https://dx.doi.org/10.1021/acs.jpclett.8b03679}

\bibitem{Ma2017}
X. Ma, Q. Liu, D. Xu, Y. Zhu, S. Kim, Y. Cui, L. Zhong, and M. Liu, ``Capillary-Force-Assisted Clean-Stamp Transfer of Two-DimensionalMaterials,'' \textit{Nano Lett.} \textbf{17}(11), 6961--6967 (2017). \url{https://doi.org/10.1021/acs.nanolett.7b03449}

\bibitem{Weber2014}
P. Weber, J. G\"{u}ttinger, I. Tsioutsios, D. E. Chang, and A. Bachtold, ``Coupling Graphene Mechanical Resonators to Superconducting Microwave Cavities,'' \textit{Nano Lett.} \textbf{14}(5), 2854--2860 (2014). \url{https://doi.org/10.1021/nl500879k}

\bibitem{Nelson2013}
A. Ko\v{s}mrlj and D. R. Nelson, ``Mechanical properties of warped membranes,'' \textit{Phys. Rev. E} \textbf{88}(1), 012136 (2013). \url{https://doi.org/10.1103/PhysRevE.88.012136}

\bibitem{Gornyi2017}
I. V. Gornyi, V. Yu Kachorovskii, and A. D. Mirlin, ``Anomalous Hooke’s law in disordered graphene,'' \textit{2D Mater.} \textbf{4}, 011003 (2017). \url{https://doi.org/10.1088/2053-1583/4/1/011003}

\bibitem{RuizVargas2011}
C. S. Ruiz-Vargas, H. L. Zhuang, P. Y. Huang, A. M. van der Zande, S. Garg, P. L. McEuen, D. A. Muller, R. G. Hennig, and J. Park, ``Softened Elastic Response and Unzipping in Chemical Vapor Deposition Graphene Membranes,'' \textit{Nano Lett.} \textbf{11}(6), 2259--2263 (2011). \url{https://doi.org/10.1021/nl200429f}

\bibitem{Nicholl2017}
R. J. T. Nicholl, N. V. Lavrik, I. Vlassiouk, B. R. Srijanto, and K. I. Bolotin, "Hidden Area and Mechanical Nonlinearities in Freestanding Graphene," \textit{Phys. Rev. Lett.} \textbf{118}(26), 266101 (2017). \url{https://doi.org/10.1103/PhysRevLett.118.266101}

\bibitem{Frank2007}
I. W. Frank, D. M. Tanenbaum, A. M. van der Zande, and P. L. McEuen, ``Mechanical properties of suspended graphene sheets,'' \textit{J. Vac. Sci. Technol. B} \textbf{25}, 2558--2561 (2007). \url{https://doi.org/10.1116/1.2789446}

\bibitem{Lee2008}
C. Lee, X. Wei, J. W. Kysar, and J. Hone, ``Measurement of the Elastic Properties and Intrinsic Strength of Monolayer Graphene,'' \textit{Science} \textbf{321}(5887), 385--388 (2008). \url{https://doi.org/10.1126/science.1157996}

\bibitem{Kang2016}
P. Kang, M. C. Wang, P. M. Knapp, and S. W. Nam, ``Crumpled Graphene Photodetector with Enhanced, Strain-Tunable, and Wavelength-Selective Photoresponsivity,'' \textit{Adv. Mater.} \textbf{28}, 4639--4645 (2016). \url{https://doi.org/10.1002/adma.201600482}

\bibitem{Chin2021}
H.-T. Chin, J. Klimes, I.-F. Hu, D.-R. Chen, H.-T. Nguyen, T.-W. Chen, S.-W. Ma, M. Hofmann, C.-T. Liang, and Y.-P. Hsieh, ``Ferroelectric 2D ice under graphene confinement'', \textit{Nat. Commun.} \textbf{12}:6291 (2021). \url{https://doi.org/10.1038/s41467-021-26589-x}

\bibitem{Moser2008}
J. Moser, A. Verdaguer, D. Jimenez, A. Barreiro, and A. Bachtold, ``The environment of graphene probed by electrostatic force microscopy,'' \textit{Appl. Phys. Lett.} \textbf{92}(12), 123507 (2008). \url{https://doi.org/10.1063/1.2898501}

\bibitem{Dorgan2013}
V. E. Dorgan, A. Behnam, H. J. Conley, K. I. Bolotin, and E. Pop, ``High-Field Electrical and Thermal Transport in Suspended Graphene,'' \textit{Nano Lett.} \textbf{13}(10), 4581--4586 (2013). \url{https://doi.org/10.1021/nl400197w}

\bibitem{Ye2018}
F. Ye, J. Lee, and P. X.-L. Feng, ``Electrothermally Tunable Graphene Resonators Operating at Very High Temperature up to 1200 K,'' \textit{Nano Lett.} \textbf{18}(3), 1678--1685 (2018). \url{https://doi.org/10.1021/acs.nanolett.7b04685}

\bibitem{Cao2020}
K. Cao, S. Feng, Y. Han, L. Gao, T. Hue Ly, Z. Xu, and Y. Lu, ``Elastic straining of free-standing monolayer graphene,'' \textit{Nat. Commun.} \textbf{11}, 284 (2020). \url{https://doi.org/10.1038/s41467-019-14130-0}

\bibitem{Jaddi2024}
S. Jaddi, M. Wasil Malik, B. Wang, N. M. Pugno, Y. Zeng, M. Coulombier, J.-P. Raskin, and T. Pardoen, ``Definitive engineering strength and fracture toughness of graphene through on-chip nanomechanics,'' \textit{Nat. Commun.} \textbf{15}, 5863 (2024). \url{https://doi.org/10.1038/s41467-024-49426-3}


\end{thebibliography}
\end{document}